\documentclass[aps,prl,twocolumn,10pt,showpacs,superscriptaddress]{revtex4}

\usepackage{graphicx}
\usepackage{amsmath,units}
\usepackage{mathrsfs}
\usepackage{color} 

\newcommand{\cF}{\mathscr{F}}
\newcommand{\eg}{{e.g., }}
\newcommand{\half}{\frac{1}{2}}
\newcommand{\ie}{{i.e., }}
\newcommand{\kT}{k_{\rm B}T}
\newcommand{\MSDa}{{\rm MSD}^{\rm 1P}}
\newcommand{\MSDb}{{\rm MSD}^{\rm 2P}}

\newcommand{\qhat}{\hat{\bf q}}

\newcommand{\vecF}{{\bf F}}
\newcommand{\vecm}{{\bf m}}
\newcommand{\vecq}{{\bf q}}
\newcommand{\vecr}{{\bf r}}
\newcommand{\vecu}{{\bf u}}
\newcommand{\vecU}{{\bf U}}
\newcommand{\vecv}{{\bf v}}
\newcommand{\rhat}{\hat{\bf r}}

\begin{document}

%============================================================
\title{Viscoelastic Response of a Complex Fluid at Intermediate Distances } 
%============================================================

\author{Adar Sonn-Segev}
%\email{adarsonn@tau.ac.il}

\affiliation{Raymond \& Beverly Sackler School of Chemistry, Tel Aviv
  University, Tel Aviv 6997801, Israel}

\author{Anne Bernheim-Groswasser}
%\email{bernheim@bgu.ac.il} 

\affiliation{Department of Chemical Engineering, Ilse Kats Institute
  for Nanoscale Science and Technology, Ben Gurion University of the
  Negev, Beer-Sheva 84105, Israel}

\author{Haim Diamant}
%\email{hdiamant@tau.ac.il}

\affiliation{Raymond \& Beverly Sackler School of Chemistry, Tel Aviv
  University, Tel Aviv 6997801, Israel}

\author{Yael Roichman}
\email{roichman@tau.ac.il}

\affiliation{Raymond \& Beverly Sackler School of Chemistry, Tel Aviv
  University, Tel Aviv 6997801, Israel}

\date{\today}

\begin{abstract}
The viscoelastic response of complex fluids is length- and time-scale dependent, encoding information on intrinsic dynamic correlations and mesoscopic structure. We study the length scale above which bulk viscoelasticity sets in, and the material response that precedes it at shorter distances. We show that the crossover between these two regimes may appear at a surprisingly large distance. We generalize the framework of microrheology to include both regimes and apply it to F-actin networks, thereby extracting their dynamic correlation length from their bulk and local viscoelastic properties.
\end{abstract}


\maketitle
%------------------------------------------------

Most fluids in nature and industry are complex, or structured
\cite{WittenBook}, in the sense that they include mesoscopic elements
in between the molecular and macroscopic scales. For example, in
suspensions, micron-scale solid particles are dispersed in a
molecular fluid, and in polymer gels the polymer chains form a network
embedded within a molecular solvent. Consequently, the response of
complex fluids to stress is characterized by intermediate length and
time scales.

The bulk viscoelastic response of such materials is commonly measured
using macrorheology \cite{LarsonBook}. Similar information, for a
wider frequency range and smaller material quantity, can be extracted
from microrheology by following the motions of embedded tracer
particles
\cite{Mason1995,Mason1997,Gittes97,Crocker2000,Squires2010,LevineLubensky}. In
one-point (1P) microrheology \cite{Mason1995,Mason1997,Gittes97} the
thermal fluctuations of a single particle are used to infer the
viscoelastic properties of the medium via a generalized
Stokes-Einstein relation (GSER). It has been found that this
measurement is affected by the local environment of the tracer
particle \cite{Chen2003,starrs2003}, and thus may fail to reproduce the
material's bulk response. Two-point (2P) microrheology
\cite{Crocker2000} overcomes this obstacle by tracking the correlated
motions of particle pairs as a function of their separation. 2P
measurements have focused on asymptotically large separations, where
the pair correlation has a universal form due to momentum
conservation.
%This procedure yields reliable values for the bulk viscoelastic
%moduli, yet fails to provide spatial information such as the
%material's intrinsic lengthscales.

The current work addresses two questions: i) Beyond what length scale does the bulk viscoelastic behavior emerge? ii) What is the material response at smaller length scales? We find that the leading correction to the
asymptotic behavior at large distances, referred to hereafter as the subdominant response, 
may be unexpectedly large, causing the bulk response to set in at surprisingly large distances. 
The physical origin of the subdominant response, which is unique to complex fluids,
is different from that of the asymptotic one. It is related as well to
a conservation law (of fluid mass rather than momentum), resulting in
a generic system-independent form. The study of this
distinctive regime leads to a more complete description of the
complex-fluid response.
 
We first derive the generic form of the subdominant response and,
subsequently, confirm the general predictions in a specific
theoretical example, the two-fluid model of polymer gels
\cite{deGennes1976,Milner1993,LevineLubensky}. Extending the framework
of microrheology to include the subdominant term, we validate its
significant effect in a model experimental system, entangled F-actin
networks of various concentrations.
  
We set the stage by recalling the classical Stokes problem of a rigid
sphere of radius $a$, driven by a steady force $\vecF$ through an
incompressible fluid of viscosity $\eta$ \cite{hydro}. The fluid
velocity at position $\vecr$ away from the sphere's center is given by
$\vecv(\vecr)=\vecv_1+\vecv_2$, with $v_{1\alpha}=(8\pi\eta
r)^{-1}(\delta_{\alpha\beta}+\rhat_\alpha\rhat_\beta)F_\beta$ and
$v_{2\alpha}=a^2(24\pi\eta
r^3)^{-1}(\delta_{\alpha\beta}-3\rhat_\alpha\rhat_\beta)F_\beta$,
where Greek indices denote the coordinates $(x,y,z)$, and repeated
indices are summed over. The dominant term at large distances,
$\vecv_1$, is the flow due to a force monopole $\vecF$. Its $r^{-1}$
decay is dictated by momentum conservation, ensuring that the
integrated momentum flux (proportional to $\nabla\vecv_1\sim r^{-2}$)
through any closed surface around the sphere remain fixed. This
dominant response can be decomposed into longitudinal and transverse
components (force and velocity parallel and perpendicular to $\vecr$,
respectively), $v_{1\parallel}=(4\pi\eta r)^{-1}F_\parallel$,
$v_{1\perp}=(8\pi\eta r)^{-1}F_\perp$, both of which are
positive. Turning to the subdominant $\vecv_2$, we point out the
largely overlooked fact that it is actually made of two contributions,
having the same spatial form but opposite signs and different physical
origins, $\vecv_2=\vecv_{2f}+\vecv_{2m}$. The first,
$\vecv_{2f}=3\vecv_2$,
%$v_{2f,\alpha}=a^2(8\pi\eta
%r^3)^{-1}(\delta_{\alpha\beta}-3\rhat_\alpha\rhat_\beta)F_\beta$,
is the flow due to a force quadrupole $Q_{\gamma\alpha\beta}=\half a^2
\delta_{\gamma\alpha} F_\beta$. We focus our attention on the opposite
contribution, $\vecv_{2m}=-2\vecv_2$.
%$v_{2m,\alpha}=-a^2(12\pi\eta
%r^3)^{-1}(\delta_{\alpha\beta}-3\rhat_\alpha\rhat_\beta)F_\beta$. 
It is the flow due to a mass dipole $\vecm=-[a^2/(3\eta)]\vecF=-2\pi
a^3\vecU$, created opposite to the direction of the sphere's displacement, where $\vecU=(6\pi\eta a)^{-1}\vecF$ is the sphere's velocity. The net
subdominant term introduces a negative correction to the longitudinal
response, $v_{2\parallel}=-a^2(12\pi\eta r^3)^{-1}F_\parallel$, and a
positive correction to the transverse one, $v_{2\perp}=a^2(24\pi\eta
r^3)^{-1}F_\perp$. Since the simple fluid has no intrinsic
lengthscale, these corrections vanish as $a\rightarrow 0$.

Now contrast the above with the case of an isotropic viscoelastic
medium \cite{ijc07}, having a frequency-dependent complex shear
modulus $G(\omega)=G'(\omega)+iG''(\omega)$ [\ie bulk shear viscosity
  $\eta_b(\omega)=G(\omega)/(-i\omega)$]. Dynamic correlations in the
medium (as measured, e.g. by dynamic scattering) decay with a
characteristic correlation length $\xi_d$, which in polymer solutions
is believed to coincide with the static mesh size $\xi_s$
\cite{deGennes1976}. Consider a sphere of radius $a$, driven through
the medium by an oscillatory force $\vecF e^{-i\omega t}$. At
sufficiently large distances the medium velocity must be dominated by
the monopolar $v_{1\alpha}=(8\pi\eta_b
r)^{-1}(\delta_{\alpha\beta}+\rhat_\alpha\rhat_\beta)F_\beta$, for the
same momentum-conservation reasons given above. This is the basis of
present 2P microrheology
\cite{Crocker2000,Squires2010,LevineLubensky}. The two subdominant
$r^{-3}$ contributions, however, become separated. Consider first the
limit $a/\xi_d\rightarrow 0$, for which the separation is
largest. (Because of the intrinsic lengthscale $\xi_d$, $v_2$ does not
vanish in this case.) The force quadrupole, $Q\sim\xi_d^2F$, creates a
flow $v_{2f}\sim\xi_d^2(\eta_b r^3)^{-1}F$, dependent (like the
monopolar $v_1$) on bulk viscosity. By contrast, the mass dipole in
this limit arises from fluid displacement at scales smaller than
$\xi_d$, where the relevant viscosity is the solvent's, $\eta$; hence
$m\sim -(\xi_d^2/\eta)F$, creating a flow $v_{2m}\sim -\xi_d^2(\eta
r^3)^{-1}F$. Thus, $v_{2m}$ is enhanced relative to $v_{2f}$ by a
factor of $\eta_b/\eta$, which is typically very large. In such a case
of a large contrast between local and bulk response, the mass-dipole
term takes over the subdominant response and changes its sign,
$v_{2\alpha}\simeq v_{2m,\alpha} \sim -\xi_d^2(\eta
r^3)^{-1}(\delta_{\alpha\beta}-3\rhat_\alpha\rhat_\beta)F_\beta$. This
has two distinctive consequences: (a) The corrections to the
longitudinal and transverse responses flip signs,
$v_{2\parallel}=\xi_d^2(\eta r^3)^{-1}F_\parallel$, $v_{2\perp}\sim
-\xi_d^2(\eta r^3)^{-1}F_\perp$. (b) The crossover to the asymptotic
$r^{-1}$ term is pushed further to a distance
$r_c\sim(\eta_b/\eta)^{1/2}\xi_d\gg\xi_d$. In the opposite limit of an
arbitrarily large sphere, $a/\xi_d\rightarrow\infty$, only bulk
properties matter, and we have $Q\sim a^2F$, $m\sim -(a^2/\eta_b)F$,
making $v_{2f}$ and $v_{2m}$ comparable again. To interpolate between
the two limits we define a local viscosity at the scale of the probe,
$\eta_\ell\equiv F/(6\pi aU)$, as determined from the sphere's
velocity \cite{eta_ell}. Additionally, dimensionless scale functions
may be introduced, satisfying $Q=a^2f(\xi_d/a)F$ and
$m=-(a^2/\eta_\ell)g(\xi_d/a)F$, such that both $f(x)$ and $g(x)$
interpolate between values $\sim 1$ for $x\ll 1$ and $\sim x^2$ for
$x\gg 1$.

We demonstrate the validity of these predictions in the two-fluid
model of a dilute polymer gel
\cite{deGennes1976,Milner1993,LevineLubensky}.  In this model an
incompressible viscous fluid with velocity field $\vecv(\vecr,t)$,
pressure field $p(\vecr,t)$, and viscosity $\eta$, is coupled to a
dilute elastic (or viscoelastic) network with displacement field
$\vecu(\vecr,t)$ and Lam\'e coefficients $\mu$ and $\lambda$ via a
mutual friction coefficient $\Gamma$ \cite{inertia}.  For a point
force acting on the fluid component, one obtains for the
fluid-velocity response in Fourier space
[$(\vecr,t)\rightarrow(\vecq,\omega)$] \cite{LevineLubensky},
\begin{equation}
  v_\alpha(\vecq,\omega) = \frac{1+(\eta_b/\eta)\xi_d^2q^2}
  {\eta_bq^2(1+\xi_d^2q^2)} (\delta_{\alpha\beta}-\qhat_\alpha\qhat_\beta)F_\beta,
\end{equation}
with $\eta_b=\eta-\mu/(i\omega)$ and
$\xi_d^2=\eta\mu/[\Gamma(\mu-i\omega\eta)]$. Inverting back from
$\vecq$ to $\vecr$ while assuming $\eta_b\gg\eta$, we get at large
distances the two predicted terms, $\vecv \simeq \vecv_1+\vecv_2$, where
\begin{equation}
  v_{1\alpha} = \frac{\delta_{\alpha\beta}+\rhat_\alpha\rhat_\beta}
     {8\pi\eta_br} F_\beta,\ 
  v_{2\alpha} =  - \frac{\xi_d^2(\delta_{\alpha\beta} - 3\rhat_\alpha\rhat_\beta)}
   {4\pi\eta r^3} F_\beta.
\label{deltaresponse}
\end{equation}
These results are for the limit $a/\xi_d\rightarrow 0$, where $\eta_\ell \rightarrow \eta$.  We have
calculated also the fluid-velocity response of this model to a forced
rigid sphere of finite radius $a$. The $\xi_d^2$ coefficient in
Eq.\ (\ref{deltaresponse}) is then modified to $a^2g(\xi_d/a)$ with
$g(x)$ given below \cite{unpublished}. The dominant response becomes
equal to the subdominant one at the distance,
\begin{equation}
  r_c = a [2(\eta_b/\eta_\ell)g(\xi_d/a)]^{1/2},\ \
  g(x) = x^2+x+1/3.
\label{rc}
\end{equation}
These expressions were obtained assuming $\eta_b/\eta_\ell\gg 1$ and
an incompressible network ($\lambda\rightarrow\infty$ or Poisson ratio
$\sigma=1/2$). A large $\eta_b/\eta_\ell$ ratio is expected, \eg for
small probes in stiff polymer networks \cite{Kilfoil}. Effects of
compressibility \cite{Levine08} are found not to change
Eq.\ (\ref{rc}) appreciably for $\sigma$ as low as $0.4$
\cite{unpublished}.

Let us summarize the three main characteristics of the subdominant
response, expected in a complex fluid with a large $\eta_b/\eta_\ell$
contrast: (a) a positive $r^{-3}$ decay of the longitudinal response;
(b) a negative transverse response; (c) a crossover to the asymptotic
response at a distance much larger than the correlation
length \cite{early_report}.

We use thermally equilibrated, homogeneous samples of entangled F-actin networks, whose rheology has been thoroughly
characterized in recent years
\cite{Crocker2000,Palmer1999,Gardel2003,Liu2006,Atakhorrami2013}. It
is well established that 1P microrheology underestimates the bulk
viscoelastic moduli of these networks, whereas a more accurate
measurement is obtained by 2P microrheology
\cite{Crocker2000,Gardel2003,Liu2006,Atakhorrami2013}.  The large
contrast between the bulk and local moduli makes these networks a good
model system for checking the aforementioned predictions. F-actin
networks have the additional benefit of an easy control over the
network's mesh size, $\xi_s=0.3/\sqrt{c_A}$, determined by the monomer
concentration $c_A$ ($c_A$ in mg/ml and $\xi_s$ in $\mu$m)
\cite{Schmidt1989}.

Entangled F-actin networks were polymerized from purified monomer
G-actin \cite{supplemental} in the presence of passivated
polystyrene colloidal particles of radii $a=0.245$ and $0.55$ $\mu$m
(Invitrogen) \cite{supplemental}. We set the average filament length
to be $\approx 13$~$\mu$m by addition of capping protein. The actin
concentrations were $c_A=0.46$--$2~\unit{mg/ml}$, corresponding to
$\xi_s=0.44$--$0.21$ $\mu$m, respectively. Immediately after
polymerization the sample was loaded into a glass cell, previously
coated with methoxy-terminated polyethylene glycol to prevent binding of the network
to the glass \cite{supplemental}. After equilibration for 30 min at
room temperature, samples were fluorescently imaged at
$\lambda=605~\unit{nm}$. Tracer particle motion from approximately
$8\times10^5$ frames per sample was
recorded at a frame rate of $70$ Hz and tracked with accuracy of at least $13$~nm \cite{Crocker1996}.

%\vspace{1em}
\begin{figure}[tbh]
%\centering
\includegraphics[scale=0.32]{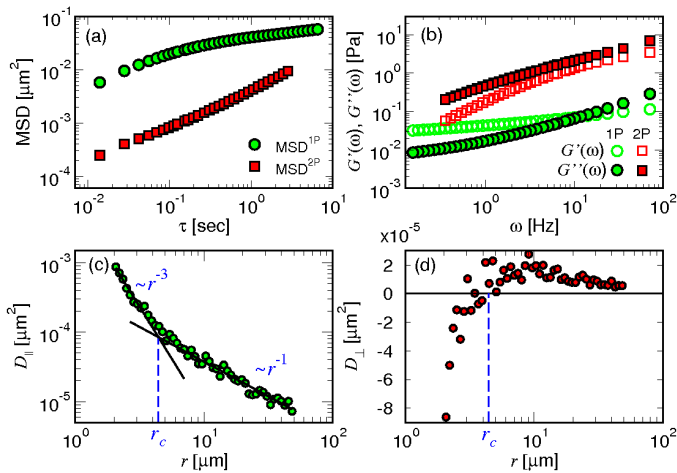}
\caption{Microrheology of entangled F-actin networks. (a) $\MSDa$
  (green) and $\MSDb$ (red) as a function of lag time, for $\xi_s=0.3$
  $\mu$m and $a=0.245$ $\mu$m. (b) The storage modulus $G'(\omega)$
  (open symbols) and loss modulus $G''(\omega)$ (filled symbols),
  extracted from the $\MSDa$ (green) and $\MSDb$ (red) curves of panel
  (a). (c) Longitudinal and (d) transverse displacement correlations
  as a function of particle separation at lag time $\tau=0.014~\unit{s}$ for
  $\xi_s=0.44$ $\mu$m and $a=0.55~\mu m$. The crossover distance $r_c$
  (blue dashed line) is defined at the intersection of the fitted
  dominant ($r^{-1}$) and subdominant ($r^{-3}$) power-law decays of
  $D_\parallel$.}
\label{fig:msds}
\end{figure}

We start by characterizing the viscoelastic properties of the F-actin
networks using conventional 1P and 2P microrheology. In 1P
microrheology one measures the ensemble-averaged mean-squared
displacement of individual tracer particles along any axis $x$ as a
function of lag time $\tau$, $\MSDa(\tau) \equiv \langle \Delta
x^2(\tau)\rangle$, and extracts from it the viscoelastic moduli,
$G'(\omega)$ and $G''(\omega)$, using the GSER
\cite{Mason1995,Crocker2007,Squires2010}. In 2P microrheology one
measures the ensemble-averaged longitudinal (parallel to $\vecr$) and
transverse (perpendicular to $\vecr$) displacement correlations of
particle pairs as functions of interparticle distance $r$ and lag time
$\tau$, $ D_\parallel(r,\tau)$, $D_\perp(r,\tau)$ \cite{Crocker2000}.
At sufficiently large distances both correlations decay as $r^{-1}$,
$D_\parallel\simeq A(\tau)/r$ and $D_\perp\simeq A(\tau)/(2r)$. The
common practice is to use this asymptote to define a `two-point
mean-squared displacement', $\MSDb(\tau) \equiv 2A(\tau)/(3a)$
\cite{msd1d}, and extract from it the viscoelastic moduli using again
the GSER \cite{Crocker2000}. Figures~\ref{fig:msds} (a) and (b) show
the 1P and 2P MSD's measured in an actin network, and the moduli
extracted from them.  The measurements demonstrate the much softer
local environment probed by the 1P technique, compared to the bulk
response probed by the 2P one.  These results are in quantitative
agreement with previous studies on F-actin networks
\cite{Crocker2000,Gardel2003,Liu2006}.

A closer look at the 2P longitudinal correlation reveals a positive
$r^{-3}$ decay preceding the asymptotic $r^{-1}$ one
(Fig.~\ref{fig:msds}(c)). The crossover between the two regimes
appears at a distance $r_c=4.4~\mu$m, an order of magnitude
larger than the network mesh size $\xi_s$. For $r<r_c$ the transverse
correlation is found to be negative (Fig.~\ref{fig:msds}(d)). Thus,
the three qualitative features mentioned above for the intermediate
response are verified.

We now extend the formalism of microrheology to include the response
at intermediate distances. This has two goals: (a) to validate in more
detail the theoretical predictions; (b) to provide a quantitative
analysis to be used in future studies of other complex fluids. We
focus on the longitudinal displacement correlation,
$D_\parallel(r,\tau)$, which is stronger than the transverse one, and
apply it in the time (rather than frequency) domain to minimize data
manipulation.

The correlation can be well fitted over both large and intermediate
distances by
\begin{equation}
  D_\parallel(r,\tau) = A(\tau)/r + B(\tau)/r^3.
\label{AB}
\end{equation}
There are three directly measured quantities: $\MSDa(\tau)$; $A(\tau)$
or, equivalently, $\MSDb(\tau)$; and $B(\tau)$. We need to relate them
to the frequency-dependent coefficients appearing in
Eq.\ (\ref{deltaresponse}). At sufficiently large distances, $r\gg a$,
the 2P coupling mobility coincides with the fluid velocity response at a
distance $\vecr$ away from an applied unit force. Using
Eq.\ (\ref{deltaresponse}), we get for the longitudinal part of that
mobility, $M_\parallel(r,\omega) = (4\pi\eta_br)^{-1} +
a^2g(\xi_d/a)(2\pi\eta_\ell r^3)^{-1}$. From the
fluctuation-dissipation theorem $D_\parallel(r,\omega) =
-(2\kT/\omega^2)M_\parallel(r,\omega)$, where $\kT$ is the thermal
energy. Comparing this with Eq.\ (\ref{AB}), we identify
\begin{eqnarray}
  A(\tau) &=& [\kT/(2\pi)] \cF^{-1}\{(-\omega^2\eta_b)^{-1}\},
\label{A}\\
  B(\tau) &=& (\kT/\pi) a^2 g(\xi_d/a) \cF^{-1}\{(-\omega^2\eta_\ell)^{-1}\},
\label{B}
\end{eqnarray}
where $\cF^{-1}$ denotes the inverse Fourier transform.  

\begin{figure}[t!]
%\centering
\includegraphics[scale=0.33]{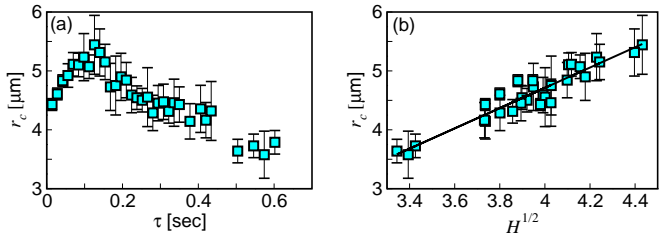}
\caption{Crossover distance as a function of (a) lag time, and (b)
  square root of $H(\tau)$, the experimental function characterizing
  the bulk to local viscosity ratio. Parameter values are $a=0.55$
  $\mu$m, $\xi_s=0.44$~$\mu$m.}
\label{fig:2Prc}
\end{figure}

\begin{figure}[t!]
\centering
\includegraphics[scale=0.33]{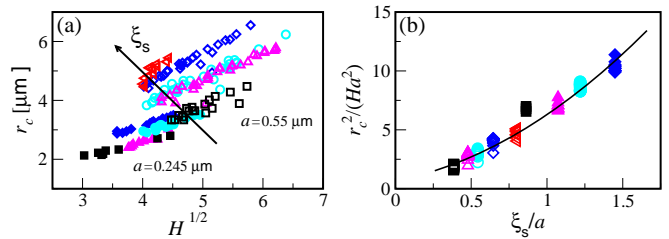}
\caption{Crossover distance for all experiments. (a) For all
  conditions $r_c$ is linear with $H^{1/2}$ and increases with either
  $\xi_s$ or $a$. (b) All experimental results fall on a master curve
  once $r_c^2$ is normalized by $Ha^2$ and presented as a function of
  $\xi_s/a$. The solid line is a fit to Eq.\ (\ref{rcH}) with the
  scale function given by Eq.\ (\ref{rc}) and $\xi_d=1.25\xi_s$. Open
  (filled) symbols correspond to $a=0.55$ ($0.245$) $\mu$m. Each color
  and symbol corresponds to a different mesh size: $\xi_s=0.21$ (black
  squares), $0.26$ (magenta triangles), $0.3$ (cyan circles), $0.35$
  (blue diamonds), and $0.44$ $\mu$m (red left triangles).}
\label{fig3}
\end{figure}

Equation (\ref{A}) merely restates the basic relation used in standard
2P microrheology to measure the bulk viscoelastic moduli. Equation
(\ref{B}) represents our extension. Its left-hand side is a directly
measurable coefficient, $B(\tau)$, while its right-hand side depends
on two dynamic characteristics of the fluid, $\eta_\ell$ and
$\xi_d$. The local response is
obtainable from the 1P measurement. According to the GSER, $\MSDa(\tau)
= [\kT/(3\pi a)]\cF^{-1}\{(-\omega^2\eta_\ell)^{-1}\}$ \cite{msd1d}.  Substitution
in Eq.\ (\ref{B}) yields a relation separating the time-dependent
observables $\MSDa(\tau)$ and $B(\tau)$ from the structural features
to be characterized, $B(\tau)/\MSDa(\tau) = 3 a^3 g(\xi_d/a)$.
Equivalently, we may examine the crossover distance,
\begin{eqnarray}
  r_c(\tau) &=& [B(\tau)/A(\tau)]^{1/2} = a[2g(\xi_d/a)]^{1/2}
  [H(\tau)]^{1/2}, \nonumber\\
  H(\tau) &\equiv& \MSDa(\tau)/\MSDb(\tau),
\label{rcH}
\end{eqnarray}
where the structural part is again decoupled from a measurable
time-dependent function, $H(\tau)$, characterizing the ratio between
the bulk and local responses.
%
%\vspace{2.3em}
In Fig.~\ref{fig:2Prc}(a) the experimentally measured $r_c$ is
plotted as a function of lag time, exhibiting a non-monotonic
dependence. Yet, by replotting $r_c$ against $[H(\tau)]^{1/2}$,
Fig.~\ref{fig:2Prc}(b), the linear dependence predicted by
Eq.\ (\ref{rcH}) is verified. 
We repeated the analysis for a set of actin networks of different
concentrations (\ie different mesh sizes) and for two different bead sizes.
Since the static and dynamic correlation lengths, $\xi_s$ and
$\xi_d$, should be comparable \cite{deGennes1976} and, 
$\xi_s$ and $a$ are comparable in our experiment, the results should
be sensitive to the interpolation function $g(\xi_d/a)$ defined in
Eq.\ (\ref{rc}). In Fig.~\ref{fig3}(a) $r_c$ for all the experiments
is plotted as a function of $H^{1/2}$. All curves are linear, as
predicted, and fall into two clusters (open and filled symbols)
corresponding to the two particle sizes. Differentiation of
Eq.\ (\ref{rc}) shows that $r_c$ should increase with either $\xi_s$
or $a$ at constant $H$, which is confirmed in Fig.~\ref{fig3}(a). For
a more quantitative validation we rescale all the measurements
according to the scheme suggested by Eq.\ (\ref{rcH})
and obtain convincing data collapse
(Fig.\ \ref{fig3}(b)). Furthermore, the resulting master curve fits well the theoretical scale function of Eq.\ (\ref{rc}) using a single free parameter\,---\,a constant ratio of order unity
between the static and dynamic lengths, $\xi_d=b\xi_s$, with $b\simeq
1.2$--$1.3$.

 One of the new insights in the current work is that the local
 viscoelastic properties of the medium affect its response over
 lengthscales much larger than the correlation length and probe
 size. Moreover, there are scenarios in which the dominant momentum
 term in the complex-fluid response is suppressed, leaving the
 subdominant mass term as the sole correlation mechanism at large
 distances. We mention three examples. (a) For a very stiff matrix, as
 in the case of a fluid embedded in a solid porous medium, the
 crossover to the asymptotic term will be pushed to arbitrarily large
 distances. (b) In a thin film of gel supported on a rigid substrate,
 the momentum term will be suppressed at distances larger than the
 film thickness, whereas the mass term will be enhanced by such
 confinement. This qualitatively accounts for the dipolar shape of the
 2P response previously reported for such a system
 \cite{sackmann2003}. (c) At sufficiently short time (high frequency)
 the diffusive momentum term is cut off beyond a certain distance
 (viscous penetration depth), whereas the mass disturbance,
 propagating via much faster compression modes, is not. All three
 scenarios obviously require further quantitative investigation.
%\vspace{2em}

Another intrinsic length scale affecting the dynamics of actin
networks is the filament length \cite{Liu2006}.  Its value in the
current work (13 $\mu$m) is much larger than $\xi_s$ and $a$.  For
shorter filament lengths there are subtle effects related to the local
environment of the probe \cite{Liu2006,unpublished}. Additional length
scales, not present in the current system, can arise from sample
heterogeneity \cite{Bonn08}.

Extracting spatio-temporal characteristics such as the dynamic
correlation length can be achieved, for example, by various dynamic
scattering techniques \cite{LarsonBook}. The intermediate response
itself, however, despite its significant effect demonstrated here,
is averaged out in such scattering measurements by virtue of the
spatial symmetry of the corresponding dipolar term.

The analysis presented here, clearly, is not restricted to actin
networks. It is applicable to any complex fluid with a sufficiently
large $\eta_b/\eta_\ell$ contrast \cite{Kilfoil}. (As `local' refers
to the scale of the probe, the contrast can be enhanced by reducing
the probe size down to $a\ll\xi_d$, whereupon the local response
becomes that of the molecular solvent.) In particular, our findings
show that bulk viscoelasticity inadequately describes micron scale
stiff biopolymer gels such as the cellular cortical network.

%%A particularly important
%system is the biological cell, whose size is comparable to the
%crossover distance measured here. Treating the cell medium as a
%viscoelstic bulk is therefore questionable, and a more detailed
%description, including the intermediate response studied here, should
%be further explored.
 
%------------------------------------------------
\begin{acknowledgments}
  We thank Rony Granek, Fred MacKintosh, and Tom Witten for helpful
  discussions. This research has been supported by the Israel Science
  Foundation (Grants No.\ 1271/08 and No.\ 8/10) and the Marie Curie
  Reintegration Grant (PIRG04-GA-2008-239378). ASS acknowledges
  funding from the Tel-Aviv University Center for Nanoscience and
  Nanotechnology.
\end{acknowledgments}

%------------------------------------------------
% References
%------------------------------------------------

%

%------------------------------------------------


\begin{thebibliography}{99}
\bibitem{eta_ell} 
The local viscosity is a function of $\xi_d/a$, with
$\eta_\ell(\xi_d/a\rightarrow\infty)\rightarrow\eta$ and
$\eta_\ell(\xi_d/a\rightarrow0)\rightarrow\eta_b$. Here we bypass this
dependence by extracting $\eta_\ell$ from 1P measurements.
\bibitem{inertia}
For simplicity we neglect inertial effects, which set in at high frequency and can be
readily added if needed \cite{LevineLubensky}.
\bibitem{early_report}
Properties (a) and (c) have been qualitatively reported for
microtubule networks \cite{Kilfoil}.
\bibitem{msd1d} 
We use the one-dimensional forms of the MSD's.
\end{thebibliography}
\end{document}